\documentclass[12pt,preprint]{aastex}
\begin{document}

\shortauthors{Stanek}

\shorttitle{AltHistAstro-ph Archive}

\title{A Modest Proposal for the Astronomical Community}

\author{Krzysztof Zbigniew Stanek\altaffilmark{1}}

\altaffiltext{1}{\small Dept.~of Astronomy, The Ohio State University, Columbus, OH 43210}

\email{kstanek@astronomy.ohio-state.edu}

\begin{abstract}

Inspired by a recent astro-ph posting, I propose a creation of an
Alternative History astro-ph archive (althistastro-ph). Such an
archive would serve as a final resting place for the various telescope
(and possibly other) proposals that were not successful. As we all
know, from both submitting proposals and also from serving on various
time allocation committees, many excellent proposals ``do not make
it''.  Creating such an AltHist archive would serve many goals,
including venting the frustration of the authors and also providing
possible amusement for the readers. These are worthy goals, but they
alone would not warrant creating such an archive. The truly useful
role of AltHistAstro-ph archive would be to match astronomers with
unappreciated ideas with other astronomers with underutilized
resources, hopefully leading in some cases to resurrection of old
proposals and resulting publications in the regular astro-ph
archive. Given the possible danger of a low signal-to-noise and
possible confusion, a creation of a separate archive seems like a good
idea, although it should be noted that low signal-to-noise is achieved
on astro-ph quite often already. Finally, I include my own excellent,
but rejected (twice), {\em HST} proposal, as an example of a potential
AltHistAstro-ph posting.

\end{abstract}

\section{Introduction}

The creation of the astro-ph archive has truly changed the way the
astronomical results are disseminated (that might also be true in the
other branches of science, but I am not familiar with those). For
example, as discussed by Schwarz \& Kennicutt (2004) and Metcalfe
(2006), ``papers that are posted to a digital preprint archive are
typically cited twice as often as papers that are not posted''
(Metcalfe 2006).

However, the current format of astro-ph is designed mostly to present
new scientific results. That is certainly a good feature, but fails to
represent a large fraction of our scientific lives devoted to writing,
mostly unsuccessful, proposals, both for research grants and also for
telescope time allocations (and by ``telescope'' I also mean other
possibly useful facilities such as X-ray, radio and other similar
instruments to gather astrophysical data).

To alleviate that shortcoming, and also being inspired by a recent
astro-ph posting, I propose a creation of an Alternative History
astro-ph archive (althistastro-ph). Such an archive would serve as a
final resting place (but see below) for the various telescope
proposals that were not successful. As we all know, from both
submitting many, many proposals and also from serving on various time
allocation committees, many excellent proposals ``do not make it''.
Creating such an AltHist archive would serve many goals, including
venting the frustration of the authors and also providing amusement
for the readers. These by themselves are worthy goals indeed, but they
alone would not warrant creating such an archive. The truly useful
role of AltHistAstro-ph archive would be to match astronomers with
unappreciated ideas with other astronomers with underutilized
resources, hopefully leading in some cases to resurrection of old
proposals and resulting publications in the regular astro-ph
archive. Given the possible danger of a low signal-to-noise and
possible confusion, a creation of a separate archive seems like a good
idea, although it should be noted that low signal-to-noise is achieved
on astro-ph quite often already.  I include my own excellent, but
rejected (twice), {\em HST} proposal, as an example of AltHistAstro-ph
posting.

\section{RR Lyr in M31}

As an example posting for the AltHistAstro-ph archive, I am including
my own {\em HST}\/ proposal that has been rejected twice by the {\em
HST} TAC. The text of the proposal has not been changed in any way
since the original submission in August 2000.

As can be seen from the included TAC comments, the ranking of my
proposal has actually decreased between the two submissions, from the
second quartile of all submitted proposals in 1999 to the third
quartile of proposals in 2000. That was despite the fact that the
second submission has incorporated the comments provided by the TAC
for the first submission.  While this seems like a paradox, it is not
surprising for somebody who has served on a {\em HST}\/ TAC: it is a
hugely oversubscribed facility, and there is a random component to all
the rankings. A cynical person could conclude that there was no point
in making any changes when submitting the second proposal, as the
memory of the previous TAC's comments is not preserved, but we should
not be cynical.

The TAC comments below are quoted {\tt verbatim}.

\begin{verbatim}

From newprop@stsci.edu Thu Dec 30 19:36:34 1999
To: kstanek@cfa.harvard.edu
Subject: Cycle 9 HST Phase I Notification Letter


Krzysztof Stanek
Smithsonian Astrophysical Observatory
MS20
60 Garden Street
Cambridge, MA  2138
United States


December 30, 1999

Dear Dr. Stanek,

We regret to inform you that following the peer review process your
proposal:

    Title:  Accurate Distance to M31 with RR Lyrae Stars

for Hubble Space Telescope Cycle 9 General Observer time has not been
approved.

Your proposal received detailed consideration by the Extra-Galactic 3
Review Panel, and final review by the STScI Director.  (The
correspondence of Science Categories to Panels for Cycle 9 can be
found on the Proposer Web page at:
http://www.stsci.edu/ftp/proposer/cycle9/cats-to-panels.html.)  Your
proposal was graded in the second quartile of proposals in your Panel.

Cycle 9 will have a duration of approximately 12 months, beginning in
July 2000.  We expect to issue the Cycle 10 Call for Proposals in June
2000, with a Phase I Deadline in early September 2000.  For your
information, 738 GO proposals requested almost 18,000 orbits in Cycle
9 , compared to the 2800 orbits available.  A total of 85 snapshot
proposals requested almost 6300 targets, compared to the 2100 targets
approved.

Comments from the peer review may be found at the end of this message.
In many cases the comments will be predominantly positive, since
oversubscription precluded the acceptance of many meritorious
proposals.  If your proposal received predominantly positive comments,
you should feel encouraged to resubmit it for a future cycle, perhaps
taking into account any suggestions made by the reviewers.

We appreciate your interest in HST, and hope that you will propose
again in the future.  There is an increasing database of HST Archival
Data, which may be a useful resource for your research and which can
include funding for U.S. proposers (for approved AR proposals).


Sincerely,

Steven Beckwith
Director


Panel Review Comments:

   Strengths: Small inconsistencies between Pop I and Pop II distance
   indicators are a nagging problem that needs to be resolved in order
   to nail down the scale /age of the Universe.  Comparisons of RRLs
   and Cepheids in M31 are fundamental. This experiment ought to
   provide a clean result on the RRL side.  The hope is to establish
   M31 at the bottom of the distance ladder which is very interesting
   (combined with the group's ground-based observations of
   Cepheids). This program will result in extremely useful archival
   data on the M31 halo.

   Weaknesses: The metallicity dependence is not the same as
   determined for our Galaxy; is there any reason to believe that a
   law of similar form ought to hold in external galaxies?  The
   estimate of the effects of crowding in the cores of the GCs seems a
   bit optimistic given how bright the GCs are.  How will they
   distinguish the halo population from that of the GCs? and how
   reliably?  The authors do not demonstrate whether or not the RRLs
   in these clusters can really provide a reliable estimate of the
   distance - for instance do these clusters have a well-defined HB or
   just a red clump?  How is this project going to improve previous
   estimates by Ajhar et al. 1991?

   Reasonableness of Resources: N/A

   Additional Comments: N/A


\end{verbatim}

\begin{verbatim}

From newprop@stsci.edu  Thu Dec 21 09:29:10 2000
To: kstanek@cfa.harvard.edu
Subject: Cycle 10 HST Phase I Notification Letter

Krzysztof Stanek
Smithsonian Astrophysical Observatory
MS20
60 Garden Street
Cambridge, MA  2138
United States


December 21, 2000

Dear Dr. Stanek,

We regret to inform you that following the peer review process your
proposal:

    Title:  Accurate Distance to M31 with RR Lyrae Stars

for Hubble Space Telescope Cycle 10 General Observer time has not been
approved.

Your proposal received detailed consideration by the Extra-Galactic 4
Review Panel, and final review by the STScI Director.  (The
correspondence of Science Categories to Panels for Cycle 10 can be
found on the Proposer Web page at:
http://www.stsci.edu/ftp/proposer/cycle10/cats-to-panels.html.)  Your
proposal was graded in the third quartile of proposals in your Panel.

Cycle 10 will have a duration of approximately 12 months, beginning in
July 2001.  We expect to issue the Cycle 11 Call for Proposals in June
2001, with a Phase I Deadline in early September 2001.  For your
information, 736 GO proposals requested over 16,000 orbits in Cycle
10, compared to the 2800 orbits available.  A total of 84 snapshot
proposals requested almost 7200 targets, compared to the 1600 targets
approved.

Comments from the peer review may be found at the end of this message.
In many cases the comments will be predominantly positive, since
oversubscription precluded the acceptance of many meritorious
proposals.  If your proposal received predominantly positive comments,
you should feel encouraged to resubmit it for a future cycle, perhaps
taking into account any suggestions made by the reviewers.

We appreciate your interest in HST, and hope that you will propose
again in the future.  There is an increasing database of HST Archival
Data, which may be a useful resource for your research and which can
lead to funding for U.S. proposers (for approved AR proposals).


Sincerely,

Steven Beckwith
Director


Panel Review Comments:

   Strengths: The distance scale is now a local problem, and it may be
   a mistake not to use HST to measure RR Lyrae in M31, since it is
   very well- suited to the problem.

   Weaknesses: It is unclear whether the resulting accuracy will be
   sufficient to help resolve the distance scale discrepencies.

   Reasonableness of Resources:  Acceptible.

   Additional Comments: N/A


\end{verbatim}

The actual proposal can be accessed at {\tt
http://www.astronomy.ohio-state.edu/\~\/kstanek/rrlyr2000.ps}.  Please
feel free to use it for whatever (non-evil) purpose you desire.

\acknowledgments

I would like to thank the participants of the morning ``Astronomy
Coffee'' at the Department of Astronomy, The Ohio State University,
for the daily and lively astro-ph discussion, one of which prompted me
to suggest the idea described in this posting. However, they do not
deserve any blame in case you do find my idea not very appealing.

\end{document}